\documentclass[12pt]{article}
\usepackage{latexsym} \usepackage{epsfig,amssymb,euscript,axodraw4j}
\usepackage{amsmath} \usepackage{bbm}

\def\be{\begin{equation}} 
\def\ee{\end{equation}}
\def\bea{\begin{eqnarray}} 
\def\eea{\end{eqnarray}}
\newcommand{\nn}{\nonumber} 
 
\newcommand{\Tr}{{\rm Tr}}

\begin{document}
\pagestyle{empty}

\begin{center}

{\LARGE{\bf Tame $D$-tadpoles in gauge mediation}}

\vspace{2cm}

{\large{Riccardo Argurio and Diego Redigolo  \\[5mm]}}

{\small{ Physique Th\'eorique et Math\'ematique\\
Universit\'e Libre de Bruxelles, C.P. 231, 1050 Brussels, Belgium\\
\medskip
and\\
\medskip
International Solvay Institutes, Brussels, Belgium}}

\vspace{2cm}

{\bf Abstract}

\vspace{1cm}

\begin{minipage}[h]{16.0cm}
We revisit models of gauge mediated supersymmetry breaking where
messenger parity is violated. Such a symmetry is usually invoked in
order to set to zero potentially dangerous hypercharge $D$-term
tadpoles. A milder hypothesis is that the $D$-tadpole vanishes only at
the first order in the gauge coupling constant. Then the next order leads to
a contribution to the sfermion masses which is of the same magnitude
as the usual radiative one. This enlarges the parameter space of gauge
mediated models. We first give a completely general characterization
of this contribution, in terms of particular three-point functions of hidden
sector current multiplet operators. We then explore the parameter
space by means of two simple weakly coupled models,
where the $D$-tadpole arising at two-loops 
has actually a mild logarithmic divergence.

\end{minipage}

\end{center}

\newpage

\setcounter{page}{1} \pagestyle{plain} \renewcommand{\thefootnote}{\arabic{footnote}} \setcounter{footnote}{0}

\section{Introduction}
Supersymmetry still remains the best understood option for physics
beyond the Standard Model (SM), in part because most of its consequences
can be calculated using perturbation theory. However, the current
trend is that the 7 TeV run (and possibly the 8 TeV run) of the LHC is gradually
excluding the simplest supersymmetric extensions of the SM (see
e.g.~\cite{Lowette:2012uh} and references therein), making it
less likely to solve the hierachy problem. With this in mind, 
it is important to be sure
that the exclusion limits do cover all the possibilities for
supersymmetric models. In other words, there might still be some
less familiar or poorly explored regions in parameter space for which
the exclusion limits are less severe. 
In the following, we will set ourselves in 
the context of gauge mediation of supersymmetry breaking (see
e.g.~\cite{Giudice:1998bp} for a review).

The purpose of these notes is to investigate how releasing the
constraint of messenger parity \cite{Dvali:1996cu,Dimopoulos:1996ig}
affects the general spectrum of soft
masses in the framework of gauge mediation. In General Gauge Mediation 
\cite{Meade:2008wd} the most general soft spectrum is computed, with
however the constraint of messenger parity so as to avoid one-loop
contributions to the sfermion masses. The latter are proportional to
the $D_Y$ tadpole, and to their respective $Y$ charge. As a
consequence, some sfermions are tachyonic, typically some sleptons,
and hence messenger parity is usually imposed in order to have
a physically acceptable spectrum. In this general context, messenger parity is
defined as a (discrete) symmetry of the hidden sector that involves
sending the hypercharge current to minus itself.

Messenger parity was introduced in
\cite{Dvali:1996cu,Dimopoulos:1996ig} in models with weakly coupled
messengers, 
precisely with the aim of avoiding this dangerous $D$-tadpole. In
\cite{Dimopoulos:1996ig} however it was also remarked that, at least
in some specific models, the messenger
parity requirement could be dropped in favour of a  milder
hypothesis. 
If the messengers are coupled to the hidden sector in a way that
depends only on their representation under the GUT $SU(5)$ group, then
$\langle D_Y \rangle$ at one loop comes out proportional to
$\Tr Y$ in the 
messenger sector. Hence since $\Tr Y=0$ for every messenger, there
is no $D$-tadpole at one loop and the sfermion masses are again two-loop
quantities. Thus eventually, in the absence of messenger parity there will be
competing contributions to these soft terms from the usual gauge loops
\cite{Dimopoulos:1996gy,Martin:1996zb} and the two-loop tadpole.

Messenger parity violation can also appear in other contexts. 
For instance,  some models of
semi-direct gauge mediation  
\cite{Seiberg:2008qj} appear to be messenger parity
violating, either because the messengers are charged under a hidden $U(1)$ 
gauge group that
develops a $D$ term \cite{Elvang:2009gk}, or because the messengers
are in a non-vectorial representation of the hidden gauge group 
\cite{Argurio:2010fn}. Such models
violate messenger parity in a way which is much different from those
of \cite{Dimopoulos:1996ig}. Our aim is to treat first messenger parity
violating theories in all generality. We will then consider some
concrete and minimal models with weakly coupled messengers in order to
explore the parameter space of such models, in the spirit of
\cite{Carpenter:2008wi,Buican:2008ws}.

\section{Messenger parity in general gauge mediation}
In General Gauge Mediation (GGM) \cite{Meade:2008wd}, the
implementation of messenger parity is 
the following.

The $D$-tadpole for $U(1)_Y$ is given by
\be
\langle D_Y \rangle = g_Y \langle J_Y \rangle,
\ee
where $J_Y$ is the bottom component of the hidden sector current supermultiplet
${\cal J}_Y$ corresponding to the global symmetry that, upon gauging,
becomes the Standard Model $U(1)_Y$.
One can impose that it vanishes on symmetry grounds postulating that
the hidden sector is invariant under a parity given by
\be
{\cal J}_Y \leftrightarrow -{\cal J}_Y. 
\label{ggmmp}
\ee
However, one could just postulate that by some other reason (perhaps a
milder assumption on the hidden sector), we have $\langle J_Y \rangle=0$
without the symmetry (\ref{ggmmp}).

\begin{figure}
\begin{center}
\includegraphics[height=0.20\textheight]{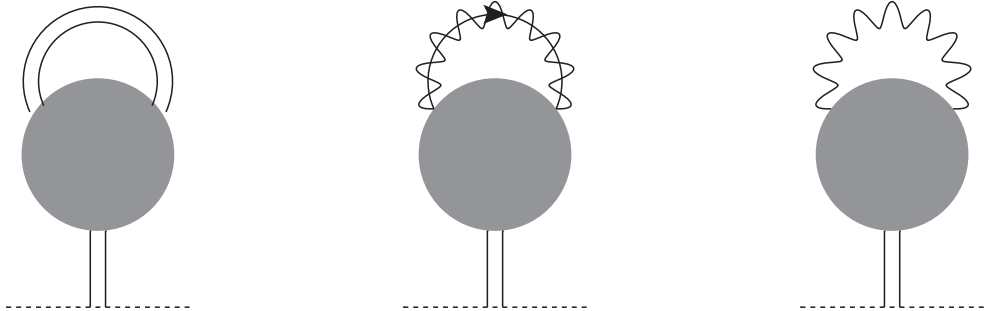}
\caption{\small Diagrammatical description of the contributions of the three point functions to the soft masses.\label{blobs}}
\end{center}
\end{figure}

Then the $D$-tadpole has a next-to-leading contribution given by
attaching a further (visible) gauge loop to the hidden sector
``blob'', as depicted in Fig.~\ref{blobs}. 
The $D$-tadpole will then be given by an expression in terms
of 3-point current correlators:
\bea
\langle D_Y \rangle &=& g_Y \sum_{i=Y,2,3} g_i^2 \int \frac{d^4p}{(2\pi)^4}  \left\{
  \langle J_Y(0)J_i(p)J_i(-p)\rangle
 +\frac{p_\mu \bar\sigma^{\mu\dot \alpha\alpha}}{p^2}
\langle J_Y(0)j_{i\alpha}(p)\bar
  j_{i\dot\alpha}(-p)\rangle \right.\nn \\ 
&& \qquad \qquad\qquad \qquad\left.
 +\frac{1}{p^2}\eta^{\mu\nu}\langle
  J_Y(0)j_{i\mu}(p)j_{i\nu}(-p) \rangle\right\}. 
\label{djjj}
\eea
The sum over $i$ runs over all the currents coupling to
$G_{SM}=SU(3)\times SU(2)\times U(1)_Y$. Note that messenger parity
would force all 3-point functions of currents to vanish, hence in
that case $\langle D_Y \rangle=0$ also at this order.

From the above expression, one can compute the contribution to the
sfermion masses:
\be
\delta m^2_{\tilde f} = g_Y Y_{\tilde f}  \langle D_Y \rangle.
\ee
We see that this contribution is of the same order as the one coming
from integrating over momentum the 2-point current correlators
\cite{Meade:2008wd}.

The above expression gives contributions of either sign, since they still depend
on each sfermion's hypercharge. However they are not parametrically
larger than the usual two-loop contribution, so that there should be a
large portion of parameter space where the soft spectrum is viable and
not finely tuned. The above contribution gives an
additional parameter on which the sfermion masses
depend. Since the vanishing of the one-loop tadpole is best motivated
in models where the hidden sector has unbroken $SU(5)$ symmetry
\cite{Dimopoulos:1996ig}, the
parameter space for sfermion 
masses in usual GGM would be reduced to one real dimension.
Here we enlarge it to two real dimensions, in a direction
``orthogonal'' to those of the original three parameters of GGM. In
particular, the sum rules $\Tr Y m^2=0 $ and $\Tr(B-L)m^2=0$ no longer
hold separately, only the combination $4\Tr Y m^2-5 \Tr(B-L)m^2=0$ is
satisfied. Such a
situation was mentioned in \cite{Meade:2008wd} but was left out of
most analyses of the GGM parameter space.

Let us be more explicit in the parametrization of the hidden sector
correlators. As usual in GGM, we write the interaction between the
hidden sector (including here the messengers) and the SSM gauge sector
by
\be
{\cal L}_\mathrm{int}  = \sum_{i=Y,2,3} g_{i} \int d^4\theta  {\cal J}_{i} V_{i} = 
\sum_{i=Y,2,3} g_{i} (J_{i} D_{i} +\lambda_{i} j_{i} + \bar \lambda_{i} \bar j_{i} + j_{i\mu} A_{i}^{\mu} ) ,
\ee
where $V_{i}$ are  vector superfields for each gauge group (which we take in the
Wess-Zumino gauge for simplicity) with coupling $g_{i}$ and
we have omitted the ${\cal O}(g_{i}^2)$ terms needed for gauge invariance. 

We have already assumed that the one-point function of $J_Y$ vanishes.
The two-point functions are parametrized as in GGM:
\bea
\langle J_{i}(p)J_{i}(-p)\rangle & = & C^{i}_0(p^2), \\
\langle j_{i\alpha}(p)\bar j_{i\dot\alpha}(-p)\rangle & = & -p_\mu
\sigma^\mu_{\alpha\dot\alpha} C^{i}_{1/2}(p^2) ,\\
\langle j_{i\mu}(p)j_{i\nu}(-p)\rangle & = & (p_\mu p_\nu - \eta_{\mu\nu}
p^2) C^{i}_1(p^2), \\
\langle j_{i\alpha}(p)j_{i\beta}(-p)\rangle & = & \epsilon_{\alpha\beta}B^{i}_{1/2}(p^2) .
\eea
We have indicated that the functions $C_s$ and $B_{1/2}$ depend on
$p^2$, but of course they will also depend on the other dimensionful
parameters of the hidden sector (such as the SUSY breaking scale(s)
and the messenger mass(es)). Note that $C_s$ are dimensionless while
in our notation $B_{1/2}$ has dimension of mass. 

We can very similarly parametrize the three-point functions needed for
the evaluation of the $D$-tadpole in (\ref{djjj}). Since all we need
are the correlators with the $J_Y$ insertion at zero momentum, the
Lorentz structure is all similar to the one of the two-point functions:
\bea
\langle J_Y(0)J_{i}(p)J_{i}(-p)\rangle & = & E^{i}_0(p^2), \\
\langle J_Y(0) j_{i\alpha}(p)\bar j_{i\dot\alpha}(-p)\rangle & = & -p_\mu
\sigma^\mu_{\alpha\dot\alpha} E^{i}_{1/2}(p^2) ,\\
\langle J_Y(0) j_{i\mu}(p)j_{i\nu}(-p)\rangle & = & (p_\mu p_\nu - \eta_{\mu\nu}
p^2) E^{i}_1(p^2).
\eea
We do not need the three-point function with two fermionic currents of
the same chirality, since it would contribute only at three loops to
$\langle D_Y \rangle$. 

Note that the $E_s$ have dimension $M^{-2}$ and 
all obviously vanish in the SUSY limit, since
there are no cubic couplings involving the $D$-term to renormalize in
the Lagrangian of a vector superfield.

Using the above expressions, the $D$-tadpole rewrites as
\be
\langle D_{Y}\rangle = \sum_{i=Y,2,3}\frac{g_{Y}g_{i}^2}{(4\pi)^2}
\int dp^2 p^2 \left(E_0^{i}(p^2) -2
E_{1/2}^{i}(p^2) + 3 E_1^{i}(p^2) \right).\label{sume}
\ee
This two-loop $D$-tadpole will give a contribution to the sfermion
masses that we  write 
\be
m^2_{\tilde f (D)} = g_{Y} Y_{\tilde f} \langle D_Y\rangle .\label{msfd}
\ee
This is to be compared with the two-loop contribution coming from the
two-point functions inserted in the usual SSM gauge radiative
corrections to the sfermion propagators:
\be
m^2_{\tilde f (C)} = -\sum_{i=Y,2,3}\frac{g_{i}^4c^{i}_{\tilde{f}}}{(4\pi)^2} \int dp^2 \left(C_0^{i}(p^2) -4
C_{1/2}^{i}(p^2) + 3 C_1^{i}(p^2) \right). \label{sumc}
\ee
where $c^{i}_{\tilde{f}}$ is the quadratic Casimir of the representation of the group
$G_i$ to which the sfermion $\tilde{f}$ belong. Notice the two main
differences between (\ref{sume}) and (\ref{sumc}): there is a factor of
2 instead of 4 multiplying the correlator involving fermionic
currents, and more importantly 
there is an extra factor of $p^2$ in the integration, that
compensates for the dimension of the $E_s$.

For completeness, the gaugino masses are given by the usual
expression
\be
m_{\lambda_{i}} = g_{i}^2 B^{i}_{1/2} (0).
\ee

Thus we see that the sfermion masses (at two-loops and at the
messenger scale) will depend on an additional
parameter which is given by the integral in (\ref{sume}).
There will be no tachyons in the sfermion sector as long as the sum of
the two contributions to each sfermion squared mass is not
negative. There is obviously a large portion of parameter space where
this is the case. 
 
In the following, we will consider simple models where we will explore
much of this parameter space.

\section{Messenger parity violation in simple models with explicit messengers}
In all generality, gauge mediation models with explicit messengers are
such that messenger
superfields $\Phi$ are in vectorial representations of $G_{SM}$. 
If one also wants to preserve unification, then the
messengers must belong to (split) multiplets of the GUT group
(typically $SU(5)$). 

Let the index $a$ run over {\em all} the components of every messenger
field. Then the current that couples to the hypercharge is simply
given by
\be
{\cal J}_Y = \sum_a Y_a \Phi_a \Phi_a^\dagger.
\ee
For instance, for a pair of messengers in the vectorial
representation ${\bf 5}+{\bf \bar 5}$, $\Phi=(\Phi^d_A,\Phi^{\bar L}_\alpha)$
we will get for the lowest component
\be
J_Y = -\frac{1}{3}\sum_A(| \phi^d_A|^2 - |\tilde \phi^{\bar d}_A|^2)
+\frac{1}{2}\sum_\alpha (|\phi^{\bar L}_\alpha|^2-|\tilde \phi^{L}_\alpha|^2).
\ee
Note that $\Tr Y=0$ for any $SU(5)$ multiplet. 
Messenger parity exchanges, in the above example, $\phi$
with $\tilde \phi^*$, which leads to $J_Y \leftrightarrow - J_Y$. 
More general set ups allow for parity exchanges
involving also unitary rotations among similar $SU(5)$ multiplets
\cite{Dvali:1996cu,Dimopoulos:1996ig}. This symmetry, or equivalent
versions, has always been assumed in all explicit computations of
sfermion masses, as for instance in \cite{Poppitz:1996xw,Cheung:2007es,Marques:2009yu,Dumitrescu:2010ha}.

The one-loop $D$-tadpole has a simple expression in terms of 2-point
functions of messengers. It is given by
\be
\langle D_Y \rangle = g_{Y} \int \frac{d^4p}{(2\pi)^4}\sum_a Y_a  \langle
\phi_a(p)\phi^*_a(-p)\rangle.
\label{dlo}
\ee 
If  the hidden
sector has a global $SU(5)$ unbroken symmetry (i.e. the breaking to
$G_{SM}$ is due to the coupling to the visible sector only), then all
the correlators $\langle \phi_a(p)\phi^*_a(-p)\rangle$ of messenger
fields belonging to the same $SU(5)$ multiplet must be the same. As a
consequence 
the sum over messengers will split into sums over $SU(5)$ multiplets,
and each term ends up being multiplied by $\Tr Y$. Then every term is
vanishing and $\langle D_Y \rangle =0$.

This no longer holds when we attach a further visible gauge loop to
the messenger loop. Essentially the $D_Y$-tadpole will be proportional
to $\sum_{i=Y,2,3}\Tr Y_a  c^i_a$ that does not vanish.\footnote{Here
  $ c^i_a$ is the quadratic Casimir of the representation of the group
  $G_i$ to which the components $a$ belong. It obviously differs for
  different $a$ part of the same GUT multiplet. For instance for a
  ${\bf 5}$ of $SU(5)$, we have $\Tr c^Y_a=\frac{5}{6}$, $\Tr
  c^2_a=\frac{3}{2}$, $\Tr c^3_a=4$ and $\Tr Y_ac^Y_a=\frac{5}{36}$, $\Tr
  Y_ac^2_a=\frac{3}{4}$, $\Tr Y_ac^3_a=-\frac{3}{4}$, while for a
  ${\bf 10}$ we have $\Tr c^Y_a=\frac{5}{2}$, $\Tr
  c^2_a=\frac{9}{2}$, $\Tr c^3_a=12$ and $\Tr Y_ac^Y_a=\frac{5}{36}$, $\Tr
  Y_ac^2_a=\frac{3}{4}$, $\Tr Y_ac^3_a=-\frac{3}{4}$.} 
On the other hand
messenger parity would
ensure that $\langle D_Y \rangle =0$ also at this order by cancelling
terms which have opposite $Y$ charges and otherwise identical hidden
sector dynamics.

We now proceed to compute the two-loop contribution to the
$D_Y$-tadpole in two simple models of messenger parity violation. The
first one contains only one pair of messengers as in Minimal Gauge
Mediation (MGM). 
Messenger parity is violated by taking the messengers to have opposite
charges with respect to a hidden $U(1)$, to which we then give a
spurionic $D$-term $\xi$. 
The second example is a revisiting of the model already discussed in 
\cite{Dimopoulos:1996ig}, where we have two pairs of messengers. In
order to have messenger parity violation, 
we need the messengers to have different SUSY
masses and an off-diagonal spurionic $F$-term. 
Both of these models have to be combined with an ordinary MGM-like
$F$-term in order to provide for non-zero gaugino masses and to uplift
all sfermion squared masses to positive values.

\subsection{A model with one messenger pair and a $D$-term $\xi$}
\label{modelone}
Our first example is based on the following messenger Lagrangian
\be
{\cal L}_\mathrm{mess} = \int d^4 \theta \left( \Phi^\dagger
e^{V_{SM}+V_h}\Phi + \tilde \Phi
e^{-V_{SM}-V_h}\tilde\Phi^\dagger\right) + \int d^2 \theta \left( M
\Phi\tilde\Phi+X\Phi\tilde\Phi\right) +c.c.
\ee
We have two spurions, $X$ and $V_h$. The chiral spurion is taken to be
$X=\theta^2F$ as in MGM, and provides for gaugino masses and for a
positive contribution to the sfermion squared masses at two-loops.
The real spurion will be taken to be 
\be
V_h=\theta^2\bar \theta^2 \xi
\ee
and is the one responsible for messenger parity violation. Indeed, the
quadratic potential for the messenger scalar sector becomes
\be
V_\mathrm{quad} = M^2 (|\phi|^2+|\tilde \phi|^2) + F(\phi\tilde\phi +
\phi^*\tilde\phi^*) +\xi (|\phi|^2-|\tilde \phi|^2)
\label{vquad}
\ee
and is no longer invariant under $\phi \leftrightarrow \tilde\phi^*$.

Note that $\xi$ does not break R-symmetry on its own. Hence it will
not contribute at leading order to the gaugino masses. Also,
R-symmetry and gauge invariance constrain $\xi$ to contribute to the
usual radiative (as opposed to tadpole) sfermion squared mass only at
the $\xi^4$ order. As remarked in
\cite{Nakayama:2007cf,Carpenter:2008rj}, such contributions can only
be written effectively as 
\be
{\cal L}_\mathrm{eff} \supset \frac{1}{\Lambda^6}\int d^4\theta  {\cal
  W}_h^2  \bar {\cal W}_h^2 Q Q^\dagger .
\ee
This actually also implies that the contribution to the $D_Y$-tadpole
will be of order $\xi^3$, corresponding to a term such as
\be
{\cal L}_\mathrm{eff} \supset \frac{1}{\Lambda^4}\int d^4\theta  
  \bar {\cal W}_h^2  {\cal W}_h^\alpha {\cal W}_{Y\alpha} \label{Wfour}
\ee
which is an effective FI term for $U(1)_Y$. 
There is also  a contribution of order $\xi$, actually related to
the kinetic mixing 
\be
{\cal L}_\mathrm{eff}\supset \log \Lambda \int d^2 \theta {\cal
  W}_h^\alpha {\cal W}_{Y\alpha},\label{mixing}
\ee
which vanishes because we take $\Tr Y=0$ over the messengers. Note
that  (\ref{mixing}) is naturally multiplied by
a logarithmic divergence.

In the appendix we perform the computation of the $C_s$ and the $E_s$
functions. 
Comparing the expressions that we obtain, 
we see that, apart from different group theory factors, they are simply related by deriving the $C_s$
with respect to $\xi$. This is actually simple to understand as
follows. Consider calculating the $C_s$ by sprinkling an (even) number
of insertions of $\xi$, treated as a perturbation to the
Lagrangian. Since the coupling of the messengers to $\xi$ is proportional
to the one to $D_Y$, we can always obtain all the diagrams
contributing to $E_s$ by removing systematically one $\xi$ insertion
and replacing it by the $D_Y$ external line. This process is
summarized by taking the derivative with respect to $\xi$ of $C_s$ in
order to obtain $E_s$. 

Amusingly, the exact relation turns out to be
\be
E^{i}_0-2E^{i}_{1/2}+3E^{i}_1 = -\frac{1}{2}\frac{\Tr Y_a  c^i_a}{\Tr c^i_a}\frac{\partial}{\partial \xi}
(C^{i}_0-4C^{i}_{1/2}+3C^{i}_1).\label{wsum}
\ee
Thus once we have the expression that we need to integrate over $p^2$
in order to get the radiative contribution to the sfermion masses, we
straightforwardly obtain the expression that we need to multiply by
$p^2$ and integrate over $p^2$ in order to get the tadpole
contribution to the sfermion masses. 

Since,  as noted in the appendix, the weighted sum of the $C_s$ is of
order $\xi^4$ when 
$F=0$, in the same situation the weighted sum of the $E_s$ will be of
order $\xi^3$. 
We can already make some observations. 
In order to avoid tachyonic sfermions, some $F$
has to be turned on. Then for $F$ and $\xi$ both small with respect to
$M^2$, we will have 
\be
m_{\tilde f (C)}^2 \propto \frac{F^2}{M^2}, \qquad 
m_{\tilde f (D)}^2 \propto \frac{\xi^3}{M^4}.
\ee
The boundary of the safe region of parameter space is then roughly
(note that $F\ll \xi$ in this region)
\be
\left(\frac{F}{M^2}\right)^2\gtrsim \left(\frac{\xi}{M^2}\right)^3.
\ee
Of course, besides computing the exact proportionality factors, one
should really RG flow the sfermion masses down to the EW scale and
make sure that there are no tachyons at that scale. Since this is
beyond the scope of the present work, we take the more
crude attitude of excluding parameter values that would give tachyons
at the messenger scale. 

Taking the limit of large external momentum we can compute the UV
behavior of the $C_{s}$ weighted sum which enters in the sfermions masses \eqref{sumc}:
\begin{align}
&C_{0}(p^{2})-4C_{1/2}(p^{2})+3C_{1}(p^2)\underset{\overset{p\to\infty}{}}{\simeq}-\frac{8}{(4\pi)^2p^4}\left[F^2\log
  \frac{p^2}{\Lambda^2}+F^2+\xi^2\right.\nn\\
&\qquad\left. -\left(M^4+\frac{1}{2}F^2\right)\log\left(1-\frac{F^2+\xi^2}{M^4}\right)-\frac{M^{2}}{\sqrt{F^2+\xi^2}}(3F^2+2\xi^2)\text{arctanh}\left(\frac{\sqrt{F^2+\xi^2}}{M^2}\right)\right]\ .\label{cexp}
\end{align}
Further expanding for small $F$ and $\xi$, we get
\be
C_{0}(p^{2})-4C_{1/2}(p^{2})+3C_{1}(p^2)\simeq -\frac{8F^2}{(4\pi)^2p^4}\left(\log
  \frac{p^2}{\Lambda^2}-1\right)+\frac{4\xi^2(F^2+\xi^2)}{3(4\pi)^2M^4p^4}+\dots
\ee
Using \eqref{wsum} and neglecting the group theory factors, we also
obtain
\be
E_{0}(p^{2})-2E_{1/2}(p^{2})+3E_{1}(p^2)\simeq -\frac{4\xi(F^2+2\xi^2)}{3(4\pi)^2M^4p^4}+\dots
\label{e1uv}
\ee
The first non-zero contribution at large momentum in \eqref{cexp} is of order $\mathcal{O}(1/p^4)$ leading to a
finite radiative contribution (\ref{sumc}) to the sfermion masses. 
This was expected, since the messenger sector satisfies the supertrace
relation $m_{+}^2+m_{-}^2=2M^2$ so that all the $C_{s}$ agree up to
$\mathcal{O}(1/p^2)$. 
On the other hand, the radiative contribution to the $D$-tadpole
(\ref{sume}) that we can easily derive from \eqref{wsum} will be
log-divergent. This divergence can be reabsorbed by wave function
renormalization of the superfields $\Phi$, $\tilde \Phi$ and $V_h$,
the latter being equivalent to the renormalization of the parameter $\xi$.
Note that the theorem of \cite{Fischler:1981zk} prevents $D$-tadpoles to be
generated at one-loop, and renormalized beyond one-loop, in a SUSY
preserving theory. However in a SUSY breaking theory such
renormalization is possible \cite{Girardello:1981wz}. 

The procedure of renormalization in the
messenger and hidden sectors introduces a choice of scheme that
cannot be fixed in terms of visible
sector observable quantities. Thus, 
this log-divergence in the $D$-tadpole makes the messenger parity
violating set up less 
predictive than the one where messenger parity is preserved. 

We will see that this feature is actually generic, since also in the
next model, which does not have any ``fundamental'' $D$-tadpole, such
a log-divergence will also arise.

\subsection{A model with two messenger pairs and an off-diagonal
  $F$-term}
\label{modeltwo}
Our second model is given by a theory with two messenger pairs
$\Phi_1,\tilde \Phi_1$ and $\Phi_2,\tilde \Phi_2$, and 
Lagrangian containing the  following superpotential
\be
W= M_1 \Phi_1\tilde \Phi_1 + M_2 \Phi_2\tilde \Phi_2 + 
X \Phi_1\tilde \Phi_1 + X \Phi_2\tilde \Phi_2 + X' \Phi_2\tilde
\Phi_1.
\ee
The spurions are
\be
X=\theta^2 F, \qquad X'=\theta^2 F'.
\ee
Messenger parity would attempt to rotate $\phi_1$ to either $\phi_1^*$
or $\phi_2^*$. In the first case it is violated by $F'$, while in the
second by the fact that $M_1\neq M_2$. This model was analyzed in
\cite{Dimopoulos:1996ig} and, when $F=0$, it shares many technical
similarities with the model considered in the previous section, albeit
it is slightly more complicated. 

It is easy to see that only when $F\neq0$ gaugino masses are
generated. However, in order to present analytical compact expressions, we
will restrict at first to the case $F=0$. If we parameterize the
$\tilde \phi_1$ and $\phi_2$ masses as
\be
M_1^2=M^2-\Delta, \qquad M_2^2=M^2+\Delta,
\ee
we see that the mass matrix of the $\tilde \phi_1$ and $\phi_2$
scalars is the same as the one of the previous model
(\ref{vquad}), with $\xi$ replaced by $\Delta$ and $F$ by
$F'$. Note however that in the present model, the fermions have split
masses $M_1$ and $M_2$, and there is an additional couple of SUSY
behaving scalars $\phi_1$ and $\tilde \phi_2$.

A straightforward computation, whose details we give in the Appendix, 
then gives directly our $C_s$ and $E_s$
functions. Note that in this case the $E_s$ functions are not simply
related to the $C_s$ as in the $D$-term case.

The expressions for the $C_{s}$ become exactly supersymmetric if ${F'}=0$, while for $\Delta=0$ we recover the usual form of MGM contributions to the sfermion masses. The expressions for the $E_{s}$ vanish if either
$F'=0$ or $M_1=M_2$ (which implies $\Delta=0$), that is when a
messenger parity can be defined.

We can make the following
observation, by expanding the expression for (\ref{sume}) 
in $F'$ and $\Delta$. The
leading term at large momenta goes like 
\be
E_0-2E_{1/2}+3E_1 = -\frac{2\Delta{F'}^4 }{5(4\pi)^2M^8p^4}+\dots,
\ee
yielding again a log-divergent two-loop contribution\footnote{We
  remark that this result is in disagreement with results reported in 
 \cite{Dimopoulos:1996ig}.}
to the $D$-tadpole, and hence to the sfermion squared masses. 

Note that in this model, the one-loop contribution, were it not for $\Tr
Y=0$, would have been finite and of ${\cal O}(\Delta {F'}^2)$
\cite{Dimopoulos:1996ig}. Indeed, the quadratic divergence cancels simply because of $U(1)_Y$'s anomaly freedom. On the other hand, 
the logarithmic divergence cancels at one loop for a more technical reason: working perturbatively in $F'$, we see that we need at least two insertions of $F'$ to close a messenger loop, so that we get at least 3 messenger propagators and the integral is finite. This result is in agreement with the general 1-loop analysis of \cite{Girardello:1981wz}.
At the two loop level there is no reason which can prevent
log-divergences to arise in the diagrams. For instance, still considering $F'$ as an insertion, there are three different cases in which logarithmic divergences could arise from 
the two-loop diagram: when all the insertions are in the loop with the
gauge or D-field, the latter is finite and constant at high momenta,
while the messenger loop will contain only two propagators;
conversely, if the gauge loop contains no insertions, this is nothing
else than an insertion of the log-divergent wave function
renormalization of the messengers, attached to a finite messenger
loop; lastly, there is one mixed case still log-divergent, when only
one insertion appears in the gauge loop. As in the previous model,
such divergences can be reabsorbed by wave function renormalization of
the two couples of messengers.  

Note also that the two-loop contribution at ${\cal O}({F'}^2)$ actually vanishes exactly, at all orders in $\Delta$. This is in line with the previous model, where at two loops there was no contribution of ${\cal O}(\xi)$, see \eqref{e1uv}.

We can evaluate $m_{\tilde f}^2$ using (\ref{msfd}) and (\ref{sumc}),
and compare the two contributions, both in the simplified set up with
$F=0$ and in the more realistic case of $F\neq 0$. 
In both cases, we expect the usual radiative contribution (\ref{sumc})
to be of ${\cal O}(F^2, {F'}^2)$, so that in this model the two loop
$D$-tadpole contribution is always subleading. 
Indeed, for $F=0$, we have:
\be
C_{0}(p^{2})-4C_{1/2}(p^{2})+3C_{1}(p^2)\simeq -\frac{8{F'}^2}{(4\pi)^2p^4}\left(\log
  \frac{p^2}{\Lambda^2}-1\right)-\frac{4\Delta^2{F'}^2}{3(4\pi)^2M^4p^4}+\dots
\ee
This means that there are no
tachyonic sfermions in all of the parameter space, but also that the
violation of the GGM sum rules will be very mild in this model.

\section{Messenger parity violation in GGM realizations of gaugino mediation}
As a last example, let us briefly consider a whole class of gauge
mediated models, namely those that implement gaugino mediation 
\cite{Kaplan:1999ac,Chacko:1999mi,Csaki:2001em,Cheng:2001an}.

In this class of models, the sfermion masses are suppressed with respect
to gaugino masses. This can be understood roughly as follows. 
In a GGM framework, the hidden sector current 2-point functions have
an additional cut off by a scale which is smaller than the typical
hidden sector (SUSY breaking) scale 
\cite{Green:2010ww,McGarrie:2010qr,Sudano:2010vt,Auzzi:2010mb}. 
The sfermion masses are
proportional to the momentum integral of such correlators, and are
hence suppressed. On the other hand gaugino masses are proportional to
the zero-momentum limit of the relevant correlator, and so they are
unchanged. Note that in such GGM realizations of gaugino mediation,
the sfermion masses at the messenger scale are always suppressed but
never completely vanishing. After RG flow, the sfermions acquire
positive squared masses of the order of the gaugino masses, suppressed
by a loop factor.

In the absence of messenger parity, we would like to known whether the
$D$-tadpole contribution to sfermion masses is suppressed or
not. Since the $D$-exchange in such contribution takes place at zero
momentum, the leading order $g^2$ contribution, if present, is not
suppressed. (Therefore a gaugino mediation set up is not viable for
solving problems related to tachyonic one-loop sfermion masses.)
In models where the leading order contribution vanishes, 
one should compute the $g^4$
contribution. Here the situation depends from the model one is
considering. In a deconstructed set up, 
the loop corrections to the $D$-tadpole are dominated by fields  of
the SM$'$ group (i.e. the group to which the 
SUSY breaking dynamics is directly coupled) and are hence
unsuppressed. As a result, it seems that messenger parity violating
models of gaugino mediation are not viable even if the $D$-tadpole is
generated at next-to-leading order.

\section{Conclusion}
We have considered gauge mediation models that violate messenger
parity. In such models, $D_Y$ acquires a tadpole. In a weakly coupled
realization of this scenario, we can have a vanishing tadpole at
one-loop by taking the
messengers in unsplit GUT multiplets. 
This contribution is however
non-vanishing at two-loops (i.e. adding a further SSM
gauge loop) and has
generically a mild logarithmic divergence,\footnote{We have shown that this two-loop contribution is controlled by a sum of specific three-point functions of the hidden sector currents. In order to show how a log-divergence always arises on general grounds, 
a careful study of the UV structure of such a function would be
needed. Also, it would interesting to address the issue of
renormalization in the hidden sector in general terms.
We leave this for further investigation.} hence making these models
less predictive than models with messenger parity (for which all
radiative corrections to the sfermion masses are strictly finite). 

Nevertheless, there can be large portions of parameter space where the
two-loop $D$-tadpole does not lead to unacceptable negative square
masses in the sfermion spectrum. We have confirmed this by analyzing
two simple models. Our main conclusion is  that, when scanning for
possible spectra associated with GGM, one should not forget the option
of choosing initial conditions for the RG flow that violate, even
maximally, the sum rules $\Tr Y m^2=0$ and $\Tr(B-L)m^2=0$, while obeying only
the linear combination $4\Tr Y m^2-5\Tr(B-L)m^2=0$. This could lead to
some phenomenologically relevant features in the low-energy spectrum,
such as NLSPs which are unusual for gauge mediated scenarios \cite{sma}.

\subsection*{Acknowledgments}
We would like to thank Gabriele Ferretti, Zohar Komargodski,
Alberto Mariotti and Marco Serone for useful discussions. 
The research of R.A. and D.R. is supported in part by IISN-Belgium (conventions
4.4511.06, 4.4505.86 and 4.4514.08), by the ``Communaut\'e
Fran\c{c}aise de Belgique" through the ARC program and by a ``Mandat d'Impulsion Scientifique" of the F.R.S.-FNRS. R.A. is a Research Associate of the Fonds de la Recherche Scientifique--F.N.R.S. (Belgium).

\appendix
\section{Some more details for the two models}
In this appendix we collect all the explicit formulas for the
two-point and three-point correlators in our two simple models. We
will omit all group theory factors, which can be easily reinstated. 

\subsection{Model with one messenger pair and a $D$-term $\xi$}
The model discussed in Section \ref{modelone} 
is simple enough to allow us in principle to write
an expression 
for the two-loop sfermion squared masses at all orders
in both $F$ and $\xi$. 

Let us first rewrite (\ref{vquad}) in matricial form
\be
V_\mathrm{quad} = (\begin{array}{cc} \phi^* & \tilde \phi\end{array})
\left(\begin{array}{cc} M^2+\xi & F \\ F & M^2-\xi \end{array}\right)
\left(\begin{array}{c} \phi \\ \tilde\phi^*\end{array}\right).
\label{massmatrix}
\ee

By rotating to 
\be
\left(\begin{array}{c} \phi \\ \tilde\phi^*\end{array}\right) =
\left(\begin{array}{cc} a & -b \\ b & a \end{array}\right)
\left(\begin{array}{c} \phi_+ \\ \phi_-^*\end{array}\right),
\ee
where 
\be
a= \frac{1}{N} (\xi + \sqrt{\xi^2+F^2}), \qquad b= \frac{1}{N} F,
\ee
\be
N^2 = 2 \sqrt{\xi^2+F^2}  (\xi + \sqrt{\xi^2+F^2}),
\ee
we obtain
\be
V_\mathrm{quad} = (\begin{array}{cc} \phi_+^* &  \phi_-\end{array})
\left(\begin{array}{cc} m_+^2 & 0 \\ 0 & m_-^2 \end{array}\right)
\left(\begin{array}{c} \phi_+ \\ \phi_-^*\end{array}\right),
\ee
with 
\be
m^2_\pm = M^2 \pm \sqrt{\xi^2+F^2}.
\ee

The coupling to the $D$-terms becomes
\be
{\cal L}\supset g D \frac{1}{\sqrt{\xi^2+F^2}}\left\{ \xi
(|\phi_+|^2 - |\phi_-|^2) - F (\phi_+\phi_- + \phi_+^*
\phi_-^*)\right\},\label{dphipm}
\ee
while the one to the gauginos is
\be
{\cal L}\supset g \sqrt{2}\lambda ( a\psi \phi_+^* -b \psi \phi_- -a
\tilde \psi \phi_-^* -b \tilde\psi \phi_+) + c.c.
\ee
The couplings of the scalars to the gauge bosons have the same
structure as in the unrotated basis:
\be
{\cal L}\supset i g A^\mu (\phi_+^*\partial_\mu \phi_+ -
\phi_+\partial_\mu \phi_+^* + \phi_-\partial_\mu \phi_-^*
-\phi_-^*\partial_\mu \phi_-).
\ee 

The first parameter we compute in this model is the one controlling
the gaugino masses which turns out to be
\be
m_{\lambda} = \frac{2g^{2} F}{\sqrt{\xi^2+F^2}}\int \frac{d^4k}{(2\pi)^4}
\frac{M}{k^2+M^2}\left(\frac{1}{k^2+m_-^2}-
\frac{1}{k^2+m_+^2}\right).
\ee
The evaluation of the integral is the same as in MGM, giving
eventually
\be
m_{\lambda} = \frac{2g^2}{(4\pi)^2}\frac{F}{M} g(x),
\ee
where
\be
g(x)= \frac{1}{x^2}\left( (1+x)\log (1+x)+(1-x)\log (1-x) \right),
\qquad x=\frac{\sqrt{\xi^2+F^2}}{M^2}.
\ee
We note that for $F=0$ the gaugino masses strictly vanish at one loop,
as expected from R-symmetry considerations.

Next we consider the expressions for the $C_s$ functions. For $C_0$ we have
\bea
C_0(p^2;\xi,F,M)&=& \int \frac{d^4k}{(2\pi)^4}\left\{ \frac{\xi^2}{\xi^2+F^2}
\left[ \frac{1}{k^2+m_+^2} \frac{1}{(p+k)^2+m_+^2} + 
 \frac{1}{k^2+m_-^2} \frac{1}{(p+k)^2+m_-^2}\right]\right. \nn \\ & & 
\qquad \qquad\qquad\qquad\qquad\left. + 2\frac{F^2}{\xi^2+F^2}\frac{1}{k^2+m_+^2}
\frac{1}{(p+k)^2+m_-^2}\right\}.
\eea
We have made explicit the dependence on all the scales
involved. Obviously only dimensionless ratios will eventually appear.

For $C_{1/2}$ we have
\be
p_\mu\sigma^\mu C_{1/2}(p^2;\xi,F,M) = 2\int \frac{d^4k}{(2\pi)^4} 
\frac{(p+k)_\mu \sigma^\mu }{(p+k)^2+M^2}\left\{\frac{1}{k^2+m_+^2}+
 \frac{1}{k^2+m_-^2}\right\}.
\ee

For the vectorial current correlator, proportional to $C_1$ we have
\bea
(p_\mu p_\nu-\eta_{\mu\nu}p^2)C_1(p^2;\xi,F,M) &=&\int
\frac{d^4k}{(2\pi)^4} (2k+p)_\mu (2k+p)_\nu \left[ \frac{1}{k^2+m_+^2}
  \frac{1}{(p+k)^2+m_+^2} \right. \nn \\
&& \qquad\qquad \qquad\left. +  
 \frac{1}{k^2+m_-^2} \frac{1}{(p+k)^2+m_-^2}\right] + \dots,
\eea
where in the ellipses are contained the terms derived from the
fermionic loop (needed to recover the SUSY part) and from the seagull
diagrams (needed for gauge invariance). 

In the above expressions we have a leading, logarthmically divergent
supersymmetric part. Singling it out, and using Feynman
parametrization to compute the loop integral, we have the following
compact expressions:
\bea
C_0 &=& C_0^\mathrm{SUSY} -\frac{1}{(4\pi)^2}\int_0^1dy 
\left\{
\frac{\xi^2}{\xi^2+F^2}\log\left(1-\frac{\xi^2+F^2}{[y(1-y)p^2+M^2]^2}\right) 
\right. \nn \\
&& \qquad\qquad\qquad\qquad \qquad\left. + \frac{F^2}{\xi^2+F^2} \log
\left(1-\frac{(1-2y)^{2}(\xi^2+F^2)}{[y(1-y)p^2+M^2]^2}\right) \right\},
\label{c0y} \\
C_{1/2}&=&C_0^\mathrm{SUSY}  -\frac{2}{(4\pi)^2}\int_0^1dy y
\log\left(1-\frac{y^2(\xi^2+F^2)}{[y(1-y)p^2+M^2]^2}\right),\label{chy}\\
C_{1} &=&C_0^\mathrm{SUSY} -\frac{1}{(4\pi)^2}\int_0^1dy (1-2y)^2
\log\left(1-\frac{\xi^2+F^2}{[y(1-y)p^2+M^2]^2}\right).\label{c1y}
\eea
It can be checked that for $F=0$, the leading
order in $C_0-4C_{1/2}+3C_1$ is indeed of order $\xi^4$. 

We now proceed to computing the $D$-tadpole. For the sake of
completeness, we first compute the 1-point function at 1-loop
\be
\left\langle J(0)\right\rangle=\int \frac{d^4k}{(2\pi)^4} 
\frac{\xi}{\sqrt{\xi^2+F^2}}\left\{\frac{1}{k^2+m_+^2}-
 \frac{1}{k^2+m_-^2}\right\}.
\ee
It has a divergent and a finite part
\be
\left\langle
J(0)\right\rangle=-\frac{2\xi}{(4\pi)^2}\left[\ln\frac{\Lambda^2}{M^2}-\int_{0}^{1}dy\ln\left(1-(1-2y)\frac{\sqrt{\xi^2+F^2}}{M^2}\right)\right]\ .
\ee
It is clear that the divergent part, proportional to the logarithm of
the UV cut-off scale $\Lambda$, is nothing but the renormalization
of the Fayet-Iliopoulos parameter of the hidden sector $\xi$. Integrating on the Feynman parameter the finite part we get
\begin{align}
\left\langle J(0)\right\rangle&=-\frac{2\xi}{(4\pi)^2}\left[\ln\frac{\Lambda^2}{M^2}+1\right.\notag\\
&\qquad\left.-\frac{M^2}{\sqrt{\xi^2+F^2}}\text{arctanh}\left(\frac{\sqrt{\xi^2+F^2}}{M^2}\right)-\frac{1}{2}\ln\left(1-\frac{\xi^2+F^2}{M^4}\right)\right]\ .
\label{oneptone}
\end{align}
Thanks to the fact that
the messengers are in degenerate complete GUT multiplets, the one-loop
tadpole cancels. 

The expressions for the three-point functions are then the
following. For the one involving only $D$-insertions, we have
\bea
E_0 & = & \int \frac{d^4k}{(2\pi)^4} \left\{
\frac{\xi^3}{(\xi^2+F^2)^{3/2}} \left[\frac{1}{(k^2+m_+^2)^2}
\frac{1}{(p+k)^2+m_+^2} -
\frac{1}{(k^2+m_-^2)^2}\frac{1}{(p+k)^2+m_-^2} \right] \right. \nn \\
 & &  \qquad+ \frac{\xi F^2}{(\xi^2+F^2)^{3/2}} 
\left[\frac{1}{(k^2+m_+^2)^2} \frac{1}{(p+k)^2+m_-^2} -
\frac{1}{(k^2+m_-^2)^2}\frac{1}{(p+k)^2+m_+^2} \right] \nn \\
 & &  \qquad+ \left.2\frac{\xi F^2}{(\xi^2+F^2)^{3/2}} 
\frac{1}{k^2+m_+^2}\frac{1}{k^2+m_-^2} \left[\frac{1}{(p+k)^2+m_+^2} -
\frac{1}{(p+k)^2+m_-^2} \right]\right\}.\label{ezero}
\eea
The three-point function with a $D$-insertion and two gaugino insertions
yields
\be
p_\mu\sigma^\mu E_{1/2}= 2\int \frac{d^4k}{(2\pi)^4}
\frac{\xi}{\sqrt{\xi^2+F^2}} \frac{(p+k)_\mu \sigma^\mu }{(p+k)^2+M^2}
\left\{\frac{1}{(k^2+m_+^2)^2}- \frac{1}{(k^2+m_-^2)^2}\right\},
\ee
where we have used $a^2+b^2=1$. The expression for the
three-point function $E_1$ with a $D$-insertion and two gauge boson
insertions can be similarly obtained:
\bea
(p_\mu p_\nu-\eta_{\mu\nu}p^2)E_1 &=& \int \frac{d^4k}{(2\pi)^4}
\frac{\xi}{(\xi^2+F^2)^{1/2}}(2k+p)_\mu (2k+p)_\nu \left[\frac{1}{(k^2+m_+^2)^2}
\frac{1}{(p+k)^2+m_+^2}\right. \nn \\ & &
\qquad\qquad \qquad \qquad \qquad-\left.
\frac{1}{(k^2+m_-^2)^2}\frac{1}{(p+k)^2+m_-^2} \right]+ \dots,
\eea
where again we do not write the seagull terms.

Observing the expressions above for the $C_s$ and the $E_s$, as
discussed in Section \ref{modelone} we can
derive the relation (\ref{wsum}). This is a technically welcome simplification,
because it allows us to compute the $E_s$ using for instance the
expressions (\ref{c0y})--(\ref{c1y}). The leading behavior of
\eqref{sume} is discussed in Section \ref{modelone}.

\subsection{Model with two messenger pairs and an off-diagonal
  $F$-term}

In the model of Section \ref{modeltwo}, 
we first rotate to a mass eigenstate basis the $\tilde \phi_1$ and $\phi_2$
scalars, where the masses are 
\be
m^2_\pm= M^2 \pm \sqrt{\Delta^2+{F'}^2},
\ee
and we define
\be
a^2=\frac{1}{2}\left(1+\frac{\Delta}{\delta}\right), \quad
b^2=\frac{1}{2}\left(1-\frac{\Delta}{\delta}\right), \quad
2ab=\frac{F'}{\delta},\quad \delta=\sqrt{\Delta^2+{F'}^2}.
\ee
As for the previous model, we write first the expressions for the $C_s$ functions. For $C_0$ we have
\bea
C_0&=& \int \frac{d^4k}{(2\pi)^4}\left\{\frac{1}{k^{2}+M_{1}^{2}}\frac{1}{(p+k)^2+M_{1}^2}+\frac{1}{k^{2}+M_{2}^{2}}\frac{1}{(p+k)^2+M_{2}^2}\right. \nn \\ 
& & \qquad \qquad\qquad \frac{\Delta^2}{\delta^2}
\left[ \frac{1}{k^2+m_+^2} \frac{1}{(p+k)^2+m_+^2}+ \frac{1}{k^2+m_-^2} \frac{1}{(p+k)^2+m_-^2}\right] \nn \\
 & & \qquad \qquad\qquad \left. + 2\frac{{F'}^2}{\delta^2}\frac{1}{k^2+m_+^2}
\frac{1}{(p+k)^2+m_-^2}\right\}.
\eea
For $C_{1/2}$ we have
\bea
p_\mu\sigma^\mu C_{1/2}&=& 2\int \frac{d^4k}{(2\pi)^4}(p+k)_\mu \sigma^\mu\left\{\left[\frac{1}{(p+k)^2+M_{1}^2}\frac{1}{k^2+M_{1}^2}+\frac{1}{(p+k)^2+M_{2}^2}\frac{1}{k^2+M_{2}^2}\right]\right. \nn \\ 
& & \qquad \qquad\qquad\qquad\qquad+\frac{1}{(k+p)^2+M_{1}^{2}}\left[a^2\frac{1}{k^{2}+m_{-}^2}+b^2\frac{1}{k^{2}+m_{+}^2}\right]\nn \\
& & \qquad \qquad\qquad\qquad\qquad\left. +\frac{1}{(k+p)^2+M_{2}^{2}}\left[a^2\frac{1}{k^{2}+m_{+}^2}+b^2\frac{1}{k^{2}+m_{-}^2}\right]\right\}. 
\eea
For the vectorial current correlator, proportional to $C_1$ we have
\bea
(p_\mu p_\nu-\eta_{\mu\nu}p^2)C_1&=&\int
\frac{d^4k}{(2\pi)^4} (2k+p)_\mu (2k+p)_\nu\left\{    \frac{1}{(p+k)^2+M_{1}^2}\frac{1}{k^2+M_{1}^2}           \right.\nn\\
&&\qquad \qquad\qquad+\frac{1}{(p+k)^2+M_{2}^2}\frac{1}{k^2+M_{2}^2}+ \frac{1}{k^2+m_+^2}\frac{1}{(p+k)^2+m_+^2} \nn \\
&&\qquad \qquad\qquad\left. +\frac{1}{k^2+m_-^2} \frac{1}{(p+k)^2+m_-^2}\right\} + \dots,
\eea
where again in the ellipses are contained the terms derived from the
fermionic loops and from the seagull
diagrams. 

It is easy to see that the above expressions for the $C_{s}$ are
supersymmetric when ${F'}=0$, while for $\Delta=0$ we get the usual
MGM expression. The first terms in each expression come from loops of the couple of SUSY
behaving scalars $\phi_1$ and $\tilde \phi_2$ and their corresponding fermionic partners. These contributions depend purely on $M_{1}$ and $M_{2}$ and 
are obviously supersymmetric so that they will not contribute to
sfermion masses in \eqref{sumc}.

As in the previous example, we can single out the leading, logarthmically divergent
supersymmetric part and use the Feynman
parametrization to compute the loop integral, obtaining the following
compact expressions:
\bea
C_0 &=& C_0^\mathrm{SUSY} -\frac{1}{(4\pi)^2}\int_0^1dy 
\left\{
\frac{\Delta^2}{\delta^2}\log\left(1-\frac{{F'}^2}{[y(1-y)p^2+M^2]^2-\Delta^{2}}\right) 
\right. \nn \\
&& \qquad\qquad\qquad\qquad \qquad\left. + \frac{{F'}^2}{\delta^2} \log
\left(1-\frac{(1-2y)^2\delta^{2}-\Delta^2}{[y(1-y)p^2+M^2]^2-\Delta^{2}}\right) \right\},
 \\
C_{1/2}&=&C_0^\mathrm{SUSY}  -\frac{2}{(4\pi)^2}\int_0^1dy y
\left\{a^2
\log\left(1-\frac{[\Delta+y(\delta-\Delta)]^2-\Delta^2}{[y(1-y)p^2+M^2]^2-\Delta^2}\right)\right.\nn \\
&&\qquad\qquad\qquad\qquad \qquad+\left.b^2
\log\left(1-\frac{[\Delta-y(\delta+\Delta)]^2-\Delta^2}{[y(1-y)p^2+M^2]^2-\Delta^2}\right)\right\},\\
C_{1} &=&C_0^\mathrm{SUSY} -\frac{1}{(4\pi)^2}\int_0^1dy (1-2y)^2
\log\left(1-\frac{{F'}^2}{[y(1-y)p^2+M^2]^2-\Delta^2}\right).
\eea

The 1-point function at 1-loop is given by
\be
\left\langle J(0)\right\rangle=\int \frac{d^4k}{(2\pi)^4} 
\left\{\frac{1}{k^2+M_1^2}-
 \frac{1}{k^2+M_2^2}+ \frac{\Delta}{\delta}\left[\frac{1}{k^2+m_+^2}-
 \frac{1}{k^2+m_-^2}\right]\right\}.
\ee
In this second model it is finite
\be
\left\langle
J(0)\right\rangle=-\frac{2\Delta}{(4\pi)^2}\int_{0}^{1}dy\left[\ln\left(1-(1-2y)\frac{\Delta}{M^2}\right)-\ln\left(1-(1-2y)\frac{\delta}{M^2}\right)\right]\ .
\ee
The absence of divergent terms is due to the fact that there is no explicit $D$-term in the Kahler potential at tree level. Integrating on the Feynman parameter we get
\begin{align}
\left\langle J(0)\right\rangle&=-\frac{2\Delta}{(4\pi)^2}\left[\frac{M^2}{\Delta}\text{arctanh}\left(\frac{\Delta}{M^2}\right)+\frac{1}{2}\ln\left(1-\frac{\Delta^2}{M^4}\right) \right.\notag\\
&\qquad\left.-\frac{M^2}{\sqrt{\Delta^2+{F'}^2}}\text{arctanh}\left(\frac{\sqrt{\Delta^2+{F'}^2}}{M^2}\right)-\frac{1}{2}\ln\left(1-\frac{\Delta^2+{F'}^2}{M^4}\right)\right]\ ,
\end{align}
where the second line is exactly the same contribution that we
obtained in the first model \eqref{oneptone} replacing $\Delta\leftrightarrow\xi$ and $F\leftrightarrow{F'}$.
Expanding for $\Delta$ and ${F'}$ small compared to $M^2$  we get
\be
\left\langle J(0)\right\rangle=-\frac{2\Delta}{(4\pi)^2}\left[\frac{{F'}^2}{6M^4}+\dots\right],
\ee
in agreement with \cite{Dimopoulos:1996ig}. Again, this contribution
vanishes when $\text{Tr}Y_{a}=0$.

For the $E_s$
functions, we obtain:
\bea
E_0 & = & \int \frac{d^4k}{(2\pi)^4} \left\{\frac{1}{(k^2+M_1^2)^2}
\frac{1}{(p+k)^2+M_1^2} -
\frac{1}{(k^2+M_2^2)^2}\frac{1}{(p+k)^2+M_2^2} \right. \nn \\
 & & \qquad \qquad+ 
\frac{\Delta^3}{\delta^3} \left[\frac{1}{(k^2+m_+^2)^2}
\frac{1}{(p+k)^2+m_+^2} -
\frac{1}{(k^2+m_-^2)^2}\frac{1}{(p+k)^2+m_-^2} \right] \nn \\
 & & \qquad \qquad+ \frac{\Delta {F'}^2}{\delta^3} 
\left[\frac{1}{(k^2+m_+^2)^2} \frac{1}{(p+k)^2+m_-^2} -
\frac{1}{(k^2+m_-^2)^2}\frac{1}{(p+k)^2+m_+^2} \right]\nn\\
 & & \qquad \qquad+ \left.2\frac{\Delta {F'}^2}{\delta^3} 
\frac{1}{k^2+m_+^2}\frac{1}{k^2+m_-^2} \left[\frac{1}{(p+k)^2+m_+^2} -
\frac{1}{(p+k)^2+m_-^2}\right]\right\}\ ,\label{ezero2}
\eea
\bea
p_\mu\sigma^\mu E_{1/2}&=& 2\int \frac{d^4k}{(2\pi)^4}(p+k)_\mu
\sigma^\mu \left\{ \frac{1}{(p+k)^2+M_1^2}\frac{1}{(k^2+M_1^2)^2}\right.\nn\\
& & \qquad \qquad \qquad\qquad \qquad-\frac{1}{(p+k)^2+M_2^2}\frac{1}{(k^2+M_2^2)^2} \nn \\
& &  \qquad- a^2 \frac{\Delta}{\delta}\left[
 \frac{1}{(p+k)^2+M_1^2}\frac{1}{(k^2+m_-^2)^2} -
\frac{1}{(p+k)^2+M_2^2}\frac{1}{(k^2+m_+^2)^2}\right] \nn \\
& &\qquad +b^2 \frac{\Delta}{\delta}\left[
 \frac{1}{(p+k)^2+M_1^2}\frac{1}{(k^2+m_+^2)^2} -
\frac{1}{(p+k)^2+M_2^2}\frac{1}{(k^2+m_-^2)^2}\right] \nn \\
& & \qquad \left.-2ab \frac{F'}{\delta} 
\left(  \frac{1}{(p+k)^2+M_1^2}-\frac{1}{(p+k)^2+M_2^2}\right)
\frac{1}{k^2+m_+^2}\frac{1}{k^2+m_-^2}\right\},
\eea
and finally
\bea
(p_\mu p_\nu -\eta_{\mu\nu}p^2)E_1 &=& \int \frac{d^4k}{(2\pi)^4}
(2k+p)_\mu(2k+p)_\nu\left\{\frac{1}{(p+k)^2+M_1^2}\frac{1}{(k^2+M_1^2)^2}
\right. \nn \\
& & \qquad \qquad \qquad \qquad\qquad-\frac{1}{(p+k)^2+M_2^2}\frac{1}{(k^2+M_2^2)^2}\nn \\
& & \qquad \qquad\qquad\qquad
+\frac{\Delta}{\delta}\left[\frac{1}{(p+k)^2+m_+^2}\frac{1}{(k^2+m_+^2)^2}
  \right.  \nn \\
& & \qquad \qquad\qquad \qquad\qquad\left.\left.
-\frac{1}{(p+k)^2+m_-^2}\frac{1}{(k^2+m_-^2)^2}\right]\right\}+\dots.
\eea
The expressions for the $E_{s}$ vanish both if
$F'=0$ or if $M_1=M_2$, $\Delta=0$, as expected. 
A lengthy computation leads to an
expression for the integrand of (\ref{sume}).
We will write first the contributions coming purely from loops of SUSY behaving scalars:   
\begin{align}
E_{0}\vert_{1,2} &= \frac{1}{(4\pi)^2}\int_{0}^{1} dy
\frac{\Delta}{(y(1-y)p^2+M^2)^2-\Delta^2}\ ,\\
E_{1/2}\vert_{1,2}&=\frac{1}{(4\pi)^2}\int_{0}^{1} dy
\frac{4y^2\Delta}{(y(1-y)p^2+M^2)^2-\Delta^2}\ ,\\
E_{1}\vert_{1,2}&=\frac{1}{(4\pi)^2}\int_{0}^{1} dy
\frac{(1-2y)^2\Delta}{(y(1-y)p^2+M^2)^2-\Delta^2}\ .
\end{align}
Note that for this subsector not only the $E_s$ do
not vanish but even
$(E_{0}^{i}-2E_{1/2}^{i}+3E_{1}^{i})\vert_{1,2}\neq 0$. 
Summing all the others  contributions we get
\bea
E_{0}\vert_{+,-}&=&-\frac{1}{(4\pi)^2}\frac{\Delta}{\delta^2}\left[\Delta^2\int_{0}^{1}dy\frac{1}{(y(1-y)p^2+M^2)^2-\delta^2}\right.\nn\\
&&+{F'}^2\int_{0}^{1}dy\frac{(1-2y)^2}{(y(1-y)p^2+M^2)^2-(1-2y)^2\delta^2}\notag\\
&&+\left. 4{F'}^2\int_{0}^{1}dy\int_{0}^{1-y}
  d{y'}\frac{(1-2{y'})}{(y(1-y)p^2+M^2)^2-(1-2{y'})^2\delta^2}\right]\ ,\\
E_{1/2}\vert_{+,-}&=&-\frac{1}{(4\pi)^2}\left\lbrace\frac{4\Delta}{\delta}\left[a^2\int_{0}^{1}dy\frac{(1-y)^{2}((1-y)\delta+y\Delta)}{(y(1-y)p^2+M^2)^2-((1-y)\delta+y\Delta)^2}\right.\right.\notag\\
&&+\left.b^2\int_{0}^{1}dy\frac{(1-y)^{2}((1-y)\delta-y\Delta)}{(y(1-y)p^2+M^2)^2-((1-y)\delta-y\Delta)^2}\right]\notag\\
&&+\left.\frac{4{F'}^2}{\delta^{2}}\int_{0}^{1}dy\int_{0}^{1-y}d{y'}\frac{y(1-y)\Delta}{\left(y(1-y)p^2+M^2+(1-y-2{y'})\delta\right)^2-y^2\Delta^2}\right\rbrace\ ,\\
E_{1}\vert_{+,-}&=&-\frac{1}{(4\pi)^2}\int_{0}^{1}dy\frac{(1-2y)^2\Delta}{(y(1-y)p^2+M^2)^2-\delta^2}\ .
\eea
These expressions can be further used to derive the leading behavior
of \eqref{sume}, as discussed in Section \ref{modeltwo}.

\end{document}